\documentstyle[11pt,newpasp,twoside,epsf,graphicx]{article}
\begin{document}
\title{Predicting Dust Distribution in Protoplanetary Discs}
\author{ L. Barri\`ere-Fouchet, J.-F. Gonzalez}
\affil{Centre de Recherche Astronomique de Lyon, \'Ecole Normale Sup\'erieure,
46 all\'ee d'Italie, 69007 Lyon, France}
\author{R.J. Humble, S.T. Maddison, J.R. Murray}
\affil{Centre for Astrophysics and Supercomputing, Swinburne University,
PO Box 218, Hawthorn, Victoria 3122, Australia}

\begin{abstract}
We present the results of three-dimensional numerical simulations that
include the effects of hydrodynamical forces and gas drag upon an
evolving dusty gas disk.  
We briefly describe a new parallel, two phase
numerical code based upon the smoothed particle hydrodynamics (SPH)
technique in which the gas and dust phases are represented by two
distinct types of particles.  We use the code to follow the dynamical
evolution of a population of grains in a gaseous protoplanetary disk in
order to understand the distribution of grains of different sizes within
the disk. Our ``grains'' range from metre to submillimetre in size.
\end{abstract}

\section{Introduction} 
The recent discoveries of extrasolar planets show our lack of knowledge
about their formation, a multi-stage process taking us from dust grains to
boulders to planetesimals to planetary embryos. Here we are primarily looking
at the initial phase -- from microns to metres.

In this paper we present the first three-dimensional numerical simulations 
that include the effects of hydrodynamical forces, self-gravity and gas drag 
upon an evolving dusty gas disk.  We use the Smoothed Particle Hydrodynamics
(SPH) (Gingold \& Monaghan 1977; Lucy 1977; Monaghan 1992) technique
which uses a collection of particles to approximate a fluid. 

 \section{Parallel Tree and Multi-Phase SPH}
We have written two slightly different SPH codes describing these
multiphase fluids. One was developed by Robin Humble (Humble et al. 2004), the other was
developed by Laure Barri\`ere-Fouchet starting from James Murray's code
(Murray 1996). Gas and dust phases are represented by two distinct types
of particles.  The equation of motion is given by Maddison (1998).

The codes have been extensively compared and the main differences arise from the fact
that Humble's code allows higher resolution simulations and Barri\`ere-Fouchet's code
seems to be able to simulate smaller size particles.

\section{Discussion}
For large ($10$m) and small ($\mu$m) dust sizes, the dust 
distribution is expected to stay close to the initial flared disk. Large grains 
(boulders) are weakly coupled to the gas, and if started in Keplerian motion, they
will remain there. Conversely, tiny grains are 
so strongly coupled to the gas that they are essentially co-moving (on the 
timescales we are examining here). 

The $0.1$mm to $10$cm interval is where all the (short timescale) interesting
dynamics takes place. In the r-z plot of figure 1, significant deviation from the 
initially flared disk occurs. The Epstein drag is sufficiently strong that it is able 
to efficiently remove energy from the dust, and yet not so strong 
that the dust is tightly coupled to the gas. 

\begin{figure}
\begin{center}
 \includegraphics[width=4cm,height=8cm,angle=90]{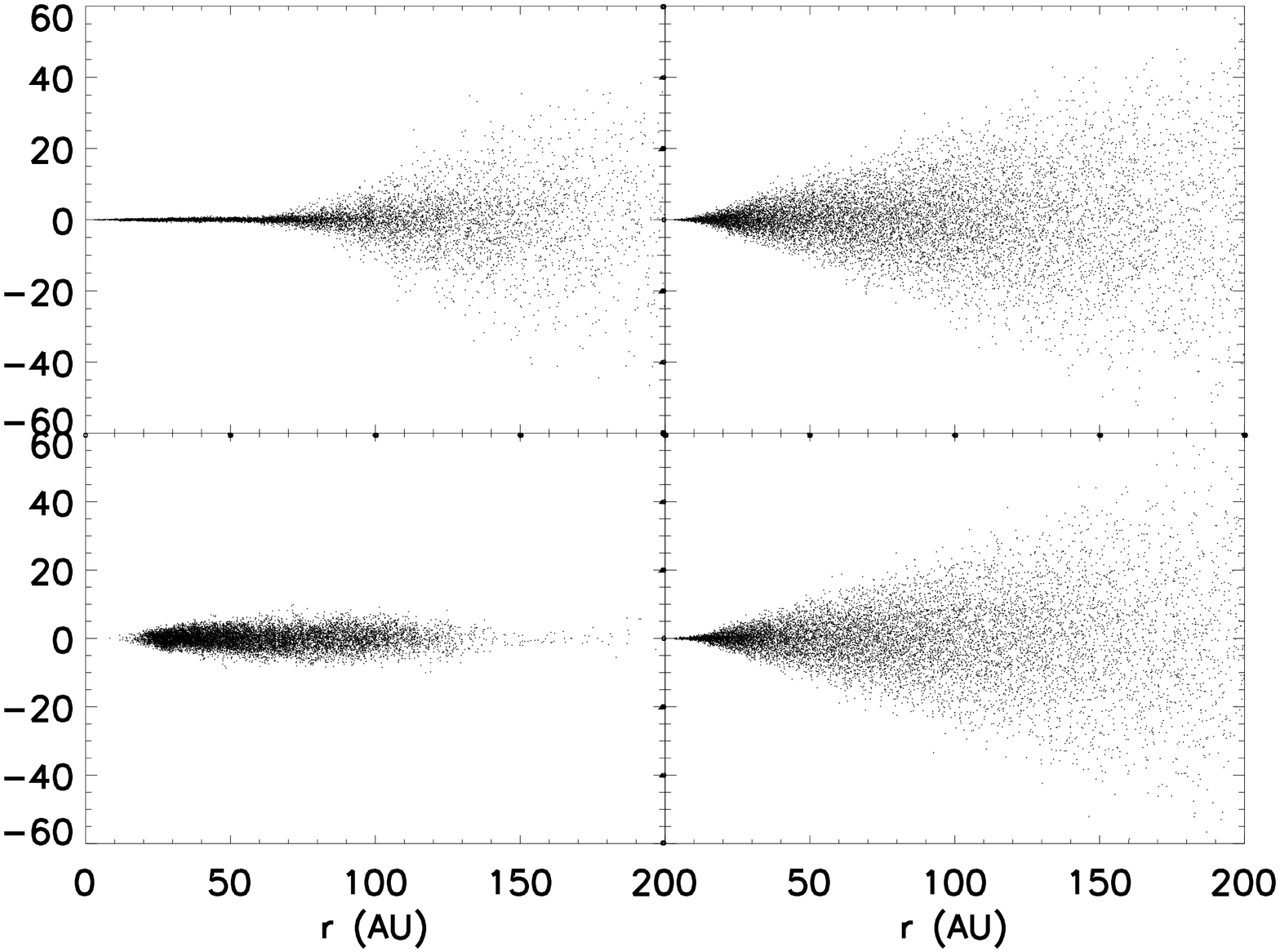}
\caption{Dust (left) and gas (right) distributions in the meridian plane of the disk.
Top: 10cm grains, bottom: 1mm grains.}
\end{center} 
\end{figure}        

\section{Summary}
We have taken the initial steps to understanding dusty disk dynamics with full 3D 
hydrodynamics.
The Lagrangian nature of the code means that it is computationally trivial to add 
empirical grain growth models, to change equations of state, and to follow the 
grain temperature 
and density histories. This means it is potentially feasible to determine not
just the location and mass of planetesimals, but also their likely composition.

\end{document}